# Exploring the Impact of Rewards on Developers' Proactive AI Accountability Behavior

*Research-in-Progress Paper*

**Long Hoang Nguyen[1]**   **Sebastian Lins[1]**   **Guangyu Du[1]**   **Ali Sunyaev[2]**
{long.nguyen, sebastian.lins, guangyu.du}@kit.edu        ali.sunyaev@tum.de
[1]Karlsruhe Institute of Technology
[2]Technical University of Munich, Campus Heilbronn

## Introduction

The widespread diffusion of Artificial Intelligence (AI)-based systems holds manifold benefits for organizations and society. Organizations, for instance, utilize AI-based systems as sales agents for enhanced business performance (Adam et al. 2022). Besides the advancing adoption of AI, scandals have revealed the risks of using AI-based systems. These risks often relate to biases, leading to discrimination against groups or individuals (Schmidt et al. 2020). A notable case of AI-induced discrimination involved an algorithm used on millions of patients that showed bias against black patients, who received less treatment as a result (Obermeyer et al. 2019). Incidents like this highlight the potential of AI-based systems to cause harm and stress the need for AI accountability. In general, AI accountability requires actors to justify their actions before a forum (Wieringa 2020).

An essential facet of AI accountability entails sanctions that punish actors and provide redress for affected parties (Busuioc 2021). The punishing nature of sanctions has manifested a negative connotation surrounding AI accountability, which is unfavorable for several reasons. First, organizations address AI accountability reactively following reputational risks (Brundage et al. 2020). This reactive approach could be problematic as organizations emphasize protecting against reputational risks over preventive processes and policies (Avin et al. 2021). Therefore, we propose a proactive approach that includes self-initiated and future-oriented actions to address risks early (Novelli et al. 2023). Second, emphasizing sanctions neglects alternative governance means like rewards. Organizations may reward AI developers to increase motivation and induce proactive behavior.

Due to the relevance of AI accountability for AI practice, several research streams emerged. First, researchers aim to define AI accountability (e.g., Wieringa 2020). Second, researchers investigate AI accountability mechanisms, including sanctions (e.g., Knowles and Richards 2021). Third, researchers examine the impact of AI accountability on stakeholders' perceptions and behaviors (e.g., Bartsch and Schmidt 2023). This study aims to bridge the second and third streams by investigating sanctions as an AI accountability mechanism and exploring rewards as an alternative means.

Despite sanctions being essential for AI accountability, research rarely elaborates on how sanctions are applied in practice (e.g., Singh et al. 2019). Research further highlights sanctions as the primary governance means for AI accountability, neglecting alternatives like rewards, despite accountability theory including both (Bovens 2007). Examining the dynamics of rewards vs. sanctions for AI accountability is promising for deeper theorizing on complementary or competing effects. Such examination would (a) shed light on alternative means for AI accountability and (b) facilitate proactive AI accountability behavior. Accordingly, we pose the research question: *How does issuing rewards vs. sanctions impact AI developers' proactive AI accountability behavior?*

This paper discusses the value of rewards vs. sanctions for AI accountability. We develop a theoretical model based on Self-Determination Theory (SDT) and Cognitive Evaluation Theory (CET). To contextualize our model with appropriate rewards and sanctions, we surveyed (AI) accountability and cybersecurity literature. Finally, we discuss our next steps and expected contributions.

# Background

## *The Need for Proactive AI Accountability Behavior*

We follow previous IS research and understand AI as "the ability of a machine to perform cognitive functions that we associate with human minds, such as perceiving, reasoning, learning, interacting with the environment, problem-solving, decision-making, and even demonstrating creativity" (Rai et al. 2019, p. iii). Due to the inherent complexity of AI, diverse challenges emerge when adopting and using AI-based systems. These challenges, among others, relate to discriminating predictions (Schmidt et al. 2020). AI accountability becomes essential to address these emerging challenges.

Accountability is used broadly across disciplines and has two meanings (Bovens 2010). First, accountability resembles a virtue referring to the willingness to act fairly and transparently. Second, accountability resembles a mechanism referring to a governance process. This process depicts "a relationship between an actor and a forum, in which the actor has the obligation to explain and to justify his or her conduct, the forum can pose questions and pass judgment, and the actor may face consequences" (Bovens 2007, p. 450). However, this definition lacks a contextualization to capture contemporary AI development processes. We align with Wieringa (2020), who contextualized AI accountability as creating an account for a socio-technical system with multiple actors. These actors must explain and justify their design, use, or decisions regarding the system.

Despite the importance of AI accountability, it has a negative connotation due to the prevalence of sanctions, which retain a penalizing nature, ultimately leading to several issues. For instance, organizations address AI accountability reactively after reputational damage (Brundage et al. 2020). This reactive approach ties AI accountability to preserving reputation over introducing preventive processes and policies (Rakova et al. 2021). Thus, there is a need for a proactive approach to foster

AI accountability. We propose proactive AI accountability behavior, which follows accountability's virtuous notion and aims to identify and rectify misconduct early on (Novelli et al. 2023).

Proactive (work) behavior emerges when employees, such as AI developers, possess the initiative to make changes for better results in their (work) environment (Parker et al. 2010). By synthesizing the literature on proactive work behavior (e.g., Crant 2000) and AI accountability (e.g., Wieringa 2020), we propose that proactive AI accountability behavior refers to employees' self-initiated and future-oriented actions that aim to justify and explain their actions and decisions related to an AI-based system. For instance, AI developers could document their decisions without external motivation. This behavior indicates that developers view AI accountability as a virtue through the willingness to act transparently (Bovens 2010) and future-oriented to prepare for later interrogations.

To induce proactive AI accountability behavior, organizations can utilize external factors, such as rewards. Prior research has shown that rewards induce behavior that individuals are not motivated to do (Deci et al. 1999). Employing rewards aligns with AI accountability definitions that explicitly mention consequences instead of sanctions (e.g., Bovens 2007; Wieringa 2020). Through rewards, organizations can induce motivation and proactivity while considering alternatives to sanctions.

### *Using Rewards to Motivate AI Developers to Take Accountability Proactively*

There are two types of motivation, namely intrinsic and extrinsic motivation. Intrinsic motivation refers to doing an activity since it is inherently enjoyable, while extrinsic motivation emerges while doing an activity for separable outcomes (Ryan and Deci 2000). These separable outcomes can motivate people to participate in activities they are not intrinsically motivated to do (Deci et al. 1999). Despite the benefits of both motivation types, evidence shows that intrinsic motivation predicts higher persistence and better performance (Cerasoli et al. 2014).

Due to the benefits of intrinsic motivation, studies have investigated its dynamics with separable outcomes. SDT by Deci and Ryan (1985) is the most prominent theory emerging from investigations. SDT proposes three psychological needs humans inherently pursue: competence, autonomy, and relatedness. SDT asserts that activities that increase these psychological needs increase intrinsic motivation (Deci and Ryan 1985). With the ongoing investigation of SDT, several sub-theories emerged. One sub-theory is CET, which emphasizes competence and autonomy and asserts that interpersonal events enhancing these needs increase intrinsic motivation (Ryan and Deci 2000).

**Theoretical Model**

Drawing on knowledge from SDT and CET, we develop a theoretical model examining the effects of rewards and sanctions on AI developers' psychological needs (Figure 1). Based on theoretical assumptions and empirical knowledge, we summarize propositions that depict the effect of rewards and sanctions on competence (P1; P3) and autonomy (P2; P4). Due to the strong empirical support for the relationship between psychological needs and intrinsic motivation, we do not derive a proposition but point to extant research (e.g., Deci et al. 1999). More importantly, we shed light on the effect of intrinsic motivation on fostering proactive AI accountability behavior of developers (P5).

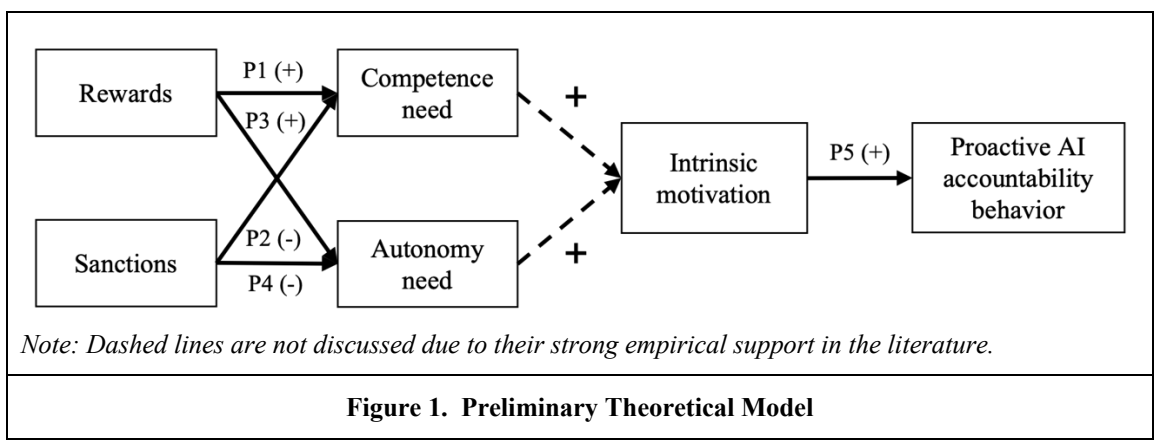

*Note: Dashed lines are not discussed due to their strong empirical support in the literature.*

**Figure 1. Preliminary Theoretical Model**

### *The Impact of Rewards and Sanctions on AI Developers' Competence*

CET proposes that interpersonal events (e.g., rewards) conveying a feeling of competence increase intrinsic motivation (Deci and Ryan 1985). To feel competent, people must feel effective and capable in their activities. For instance, employees should receive optimal challenges or encouraging feedback to feel competent while working (Ryan and Deci 2000). Prior studies (e.g., Deci 1971) showed that positive performance feedback enhances employees' competence. These findings indicate that AI accountability rewards (e.g., positive performance feedback for accountable actions) enhance AI developers' competence and intrinsic motivation. Thus, we propose that *AI accountability rewards should increase AI developers' competence and intrinsic motivation (P1)*.

Contrarily, interpersonal events that diminish competence decrease intrinsic motivation, according to CET (Deci and Ryan 1985). Prior studies examining the effect of sanctions highlight the adverse impact of discouraging feedback and threats of sanctions (e.g., Deci and Cascio 1972). Based on this, we assume that sanctioned AI developers feel less competent and intrinsically motivated (e.g., when receiving discouraging feedback after mistakes). Therefore, we propose that *AI accountability sanctions should decrease AI developers' competence and intrinsic motivation (P2)*.

### *The Impact of Rewards and Sanctions on AI Developers' Autonomy*

CET argues that intrinsic motivation further requires autonomy (Deci and Ryan 1985). Individuals must feel freedom and volition during an activity to feel autonomous. CET further suggests interpersonal events' informational vs. controlling aspects (Deci and Ryan 1985). Events that promote the informational aspect (e.g., non-pressuring rewards; Reeve 2016) increase autonomy and intrinsic motivation (Deci et al. 1999). Prior studies (e.g., Deci et al. 1981) confirm this positive impact. Considering this, AI developers should feel autonomous when rewarded in a non-pressuring way (i.e., have a choice to behave proactively). Therefore, we propose that *AI accountability rewards should increase AI developers' autonomy and intrinsic motivation (P3)*.

In contrast, CET argues that autonomy-thwarting events like pressure decrease intrinsic motivation (Deci et al. 1999). When pressured, individuals tend to feel less autonomous and intrinsically motivated (Reeve 2016). CET further highlights the informational and controlling aspects of interpersonal events. Pressuring events like threats of sanctions are generally controlling (Deci and Cascio 1972) and decrease autonomy. This decrease leads to people losing intrinsic motivation (Grolnick and Ryan 1987). Thus, we believe AI developers feel less autonomous when facing external pressure. Managers who leave AI developers with no choice but to display proactive AI accountability behavior should reduce autonomy and intrinsic motivation. Thus, we propose that *AI accountability sanctions should decrease AI developers' autonomy and intrinsic motivation (P4)*.

### The Impact of Intrinsic Motivation on Proactive AI Accountability Behavior

Research on motivation, including CET, provides a starting point by explaining how intrinsic motivation improves the persistence of behaviors (Cerasoli et al. 2014). Intrinsically motivated behaviors are inherently enjoyable and thus increase persistence (Pinder 2014). This enjoyment may also predict increased satisfaction (Van den Broeck et al. 2008). Thus, we assume that AI developers are more willing to engage in proactive accountability behavior (i.e., act fairly and transparently) when intrinsically motivated and thus satisfied. Prior research (e.g., Crant 2000) substantiates our assumption by showing that proactivity derives from intrinsic motivation. Therefore, we propose that *increased intrinsic motivation of AI developers resulting from psychological need satisfaction should induce higher levels of proactive AI accountability behavior (P5)*.

**Preliminary Findings**

Our theoretical model builds on SDT and CET and proposes that rewards increase and sanctions decrease the intrinsic motivation of AI developers. Nonetheless, contextualization of our model is required as CET stresses that the type of reward or sanction determines their impact. We performed a literature review on AI accountability to identify typical sanctions and turned to cybersecurity literature to examine promising reward mechanisms that foster proactive behavior.

We reviewed 24 studies discussing sanctions for AI accountability. Research distinguishes between punitive measures and redress (Busuioc 2021). Forums impose punitive measures when AI developers' explanations or justifications for misconduct are inadequate (Busuioc 2021). Prior research has not agreed on specific manifestations of punitive measures, which may include public criticism and fines (e.g., Singh et al. 2019). Research further argues for providing redress for victims, such as apologies and financial compensation (e.g., Fukuda-Parr and Gibbons 2021).

We identified bug bounty programs as a promising reward to foster AI accountability. Bug bounties are prevalent in cybersecurity and refer to a mechanism that allows individuals to receive compensation for identifying and reporting bugs (Brundage et al. 2020). Bug bounties helped achieve more organizational accountability (Avin et al. 2021). For AI-based systems, bugs would relate to undesirable aspects of data and algorithms (e.g., bias). Due to their potential for accountability, organizations like X introduced bias bounty challenges (Chowdhury and Williams 2021).

**Outlook and Expected Results**

This paper questions the negative connotation of AI accountability that centers around sanctions. We propose proactive AI accountability behavior, which comprises self-initiated and future-oriented actions and resembles accountability's notion as a virtue. We derived a first theoretical model highlighting how rewards foster and sanctions thwart intrinsic motivation, which predicts proactive AI accountability behavior. We uncovered typical sanctions and bug bounties as an appropriate reward mechanism by surveying (AI) accountability and cybersecurity research. Our ongoing work will contextualize our model based on these findings. We plan to test our model in a scenario-based experiment with AI developers where we manipulate the presence of rewards and sanctions.

Our research will make three core contributions. First, we propose and conceptualize proactive AI accountability behavior. Second, we expect to uncover the potential of rewards and drawbacks of sanctions for developers' proactive AI accountability behavior. In doing so, we provide novel perspectives on AI accountability and counteract the current negative connotation. Finally, we expect to contribute to CET by comparing how sanctions and rewards impact intrinsic motivation.